\documentclass[conference]{IEEEtran}

\usepackage[dvips]{graphicx}
\usepackage[cmex10]{amsmath}
\usepackage{amssymb}
\usepackage{cite}
\usepackage[all]{xy}

\begin{document}

\newtheorem{theorem}{Theorem}
\newtheorem{lemma}{Lemma}
\newtheorem{Corollary}[theorem]{Corollary}
\newtheorem{definition}{Definition}

\newcommand{\bookhskip}{\hskip 1em plus 0.5em minus 0.4em\relax}

\title{On the Capacity of Noisy Computations}
\author{
	\IEEEauthorblockN{Fran\c{c}ois Simon}
	\IEEEauthorblockA{Institut TELECOM ; Telecom SudParis ;  CITI \\ 
	9 rue Charles Fourier, 91011 EVRY Cedex, France \\
	Email: francois.simon@it-sudparis.eu}
}

\maketitle

\begin{abstract} 
This paper presents an analysis of the concept of {\em capacity}  for  {\em noisy computations}, i.e.  algorithms implemented by unreliable computing devices (e.g. noisy Turing Machines). The capacity of a noisy computation is defined and justified by companion coding theorems. Under some constraints on the encoding process, capacity is the upper bound of  {\em input rates} allowing reliable computation, i.e. decodability of noisy outputs into expected outputs.  A model of noisy computation of a perfect  function $f$ thanks to an unreliable device $F$ is given together with a model of reliable computation based on input encoding and output decoding. A coding lemma (extending the Feinstein's theorem to noisy computations), a joint source-computation coding theorem and its converse are proved. They apply if  the input source, the function $f$, the noisy device $F$ and the cascade $f^{-1}F$ induce AMS  and ergodic one-sided random processes. 
\end{abstract}

\section{Introduction and related works \label{SectionIntroduction}}

Reliable computation with unreliable devices, or in the presence of noise, has been the subject of numerous works within the vast field of fault-tolerant computing. Computation can be made reliable using information and  component/gate redundancy. Some works aim at identifying theoretical "boundaries" on the amount of necessary and/or sufficient redundancy to achieve reliability.   Recent references (see for example, \cite{Spielman96}, \cite{Gacs05},\cite{Hadjicostisverghese05}, \cite{RachlinSavage08}) continue to extend the stream opened by Von Neumann's seminal paper \cite{VonNeumann56}. These works identify bounds (e.g., depth and size of circuits) and  propose frameworks to design reliable computations mainly thanks  to gate redundancy.  These papers do not address the question of boundaries about  information redundancy. This subject, through the concept of capacity and coding theorems, has been thoroughly studied for data communication. It has not been the case for computation although the two problems (noisy computation and noisy communication) are very close: the matter is to retrieve expected values from the outputs of a random process. 

The question of whether a noisy computation possesses a capacity (or equivalently whether some coding theorems for noisy computation hold and in which cases) has been raised by P. Elias in 1958  \cite{Elias58}. There is a practical consequence in answering positively this question. This would  mean that, given a noisy implementation of an expected (e.g., Turing computable) function, it is possible to find families of efficient  input codes which asymptotically allow an almost perfect computation. Efficiency means having an input encoding rate which could remain strictly positive or arbitrarily close to a capacity when the length of the code tends to infinity. 

In a strongly constrained context (independent encoding of operands for bit-by-bit boolean operations), P. Elias obtained negative first results on the existence of a noisy computation capacity (\cite{Elias58}) . This work was deepened by Peterson and Rabin in \cite{PetersonRabin59} and by Winograd in \cite{Winograd62}. One of the major conclusions of these studies was that reliable computation with positive rate (the ratio $\frac{k}{n}$ of encoding $k$-length input block in $n$-length blocks of binary symbols) in the presence of noise  is not possible for some boolean operations (e.g., AND) under some assumptions (independent coding of operands, bijective decoding and bit-by-bit operation). This led to the conclusion that, under these assumptions, there is no capacity for such noisy operations. It is worth noting that the assumptions were made to forbid the reliable encoder and decoder to "participate" to the computation of the expected operation.

In \cite{Ahlswede84}, Ahlswede went into the subject in greater depth and made an important contribution  in characterizing  contexts  in which  a capacity for noisy computations cannot exist. It appears that the characteristics of the decoding function play a fundamental role. If the inverse of the decoding function is injective and monotonic  then weak converse theorems hold for the average and maximal error probabilities. If, in addition, the inverse of the decoding function preserves the logical AND (this implies monotonicity), then strong converses hold. These theorems state that the rate of encoding tends to $0$ when the block code length tends to infinity. The hypotheses made in \cite{Elias58,PetersonRabin59,Winograd62}, i.e., independent encoding of operands and bijective decoding, imply monotonicity of the inverse of the decoding function. On this aspect, \cite{Ahlswede84} supersedes \cite{Elias58,PetersonRabin59,Winograd62}.

Nevertheless, these negative results do not imply the absolute impossibility to identify a capacity for noisy computation. They characterize   codes, encoding and decoding processes which cannot open this ability. To  define a capacity for noisy computations, assumptions must be relaxed.

To the author's best knowledge, the first positive answer given through a definition of a capacity of a noisy computation (in fact similar to the one for a noisy channel) and a coding theorem came from  Winograd and Cowan in \cite{WinogradCowan63}.   In \cite{WinogradCowan63}, the entropy $H(X|F(X))$ of the input source conditioned by the noisy computation output is assessed as a noise measure. As it is the equivocation between the {\em noisy output} and the {\em input}, this quantity is not relevant, in full generality,  as the equivocation due to the sole noise:  it encompasses also the amount of information lost by computation. But, in a special case of noisy functions called decomposable modules, $H(X|F(X))$ actually measures the equivocation due to noise.  Decomposable modules are noisy functions  which  can be modeled by a perfect function followed by a noisy communication channel: the error probability depends on the desired output value rather than on the input value. These peculiar  functions, though noisy, make the context equivalent to  that where the reliable encoder computes and encodes the expected function result before communication through a noisy channel. Due to the restriction of considering decomposable modules, \cite{WinogradCowan63} did not completely succeed in proposing a noisy computation capacity  in a general scope (\cite{WinogradCowan63}, theorem 6.3, pages 47-48 ). 

Noisy computation capacity is also considered in reliable reconstruction of a function of sources over a multiple access channel. Much more recently, a definition of noisy computation capacity is established by Nazer and Gastpar in \cite{NazerGastpar07} and is totally consistent with the one proposed here.  Nazer and Gastpar demonstrate the possible advantages of joint source-channel coding of multiple sources over a separation-based scheme, allowing a decoder to retrieve a value which is a function of input sources. This context makes relevant the proposed distributed encoding process which perfectly performs a computation equivalent to the desired function. The encoder outputs are then transmitted through a noisy MAC to a decoder (see proofs of Theorems 1 and 2 of \cite{NazerGastpar07}). This also models a noisy computation as a perfect computation followed by a noisy transmission of the result. It can be noticed that \cite{WinogradCowan63} and \cite{NazerGastpar07} relax the assumptions of \cite{Elias58, PetersonRabin59, Winograd62, Ahlswede84} in a similar way: all goes as if the operands are jointly coded into an encoded form of the expected function result before being handled by a noisy communication channel. 

The present  paper establishes a model setting down the problem of noisy computation (section \ref{SectionNoisyComputation}), a definition of the capacity of a noisy computation with respect to an expected function and a coding lemma (section \ref{SectionCapacityofaNoisyComputation}).  A  model for reliable computation is  given, section \ref{SectionReliableComputation}. Based on this model, a joint source-computation coding theorem and its converse are stated and proved in Section \ref{SectionCodingTheorem}. This theorem aims at formally capturing  practical approaches in which reliable computation of a function $g$ is obtained thanks to a noisy apparatus $F$ computing with noise a function $f$ (e.g., a regular arithmetic addition $g$ obtained from the noisy actual circuit $F$ implementing $f$ which is an addition acting on residue encoded operands, \cite{RaoFujiwara89}). The input source, $f$, $g$, $F$ and the cascade $f^{-1}F$ are supposed to be  AMS  and ergodic one-sided random processes or channels, extending \cite{Simon10} to more general random processes and algorithms. The perfect function $f$ is assumed  {\em unary} (as is a Turing computable function). $n$-ary functions can be modeled as unary ones by concatenating $n$ input values in one "meta"-input and thus modeling a {\em joint coding} of operands. This relaxes the assumptions of  \cite{Elias58, PetersonRabin59, Winograd62, Ahlswede84}.  

\section{Model for Noisy Computation \label{SectionNoisyComputation}}

 In this section, the notations used follow \cite{Gray11} and cover countable alphabets, assumed standard (thus conditional probabilities are regular).

Let $X\equiv \{ X_i ; i\in \mathcal{I} \}$ a random process with values in  $(A^\mathcal{I}, \mathcal{B}_{A^\mathcal{I}})$ where $A$ is a countable alphabet and $\mathcal{I}$ a countable set of indexes (e.g., $\mathbb{N}$). $\mathcal{B}_{A^\mathcal{I}}$ denotes the $\sigma$-field generated by the rectangles (chapter 1, \cite{Gray11}). The random process $X$ is the source of inputs of the noisy computation. 

The noisy computing device is modeled as a random channel, i.e. a set of conditional probabilities $F \equiv \{ F_x, x \in A^\mathcal{I}\}$, taking $X$ as input and producing as an output a random process $Z\equiv \{ Z_i ; i\in \mathcal{I} \}$  on $(C^\mathcal{I}, \mathcal{B}_{C^\mathcal{I}})$ where $C$ is a countable alphabet. The hookup $P_{XZ}\equiv P_XF$ is the probability measure characterizing the actual noisy computation with an input flow represented by $X$.  From \cite{Gray11}, chapter 2, $\forall O \in \mathcal{B}_{A^\mathcal{I} \times C^\mathcal{I}}$:
$$
P_{XZ} (O) =  \int_{A^\mathcal{I}} P_{Z|X}(O_x|x) dP_X 
	 = \int_{C^\mathcal{I}} P_{X|Z}(O_z|z) dP_Z
$$
where $O_x=\left\{z\in C^\mathcal{I} / (x,z)\in O \right\}$. The probabilities $\left\{   P_{Z|X}(. |x), x\in A^\mathcal{I}\right\}$ defines the channel $X \rightarrow Z$ ($F_x \equiv P_{Z|X}(. |x)$) and $\left\{   P_{X|Z}(. |z), z\in C^\mathcal{I}\right\}$ the  "reverse" channel $F^{-1}$.
$P_{XZ}$ fully determines the channels $F$ and $F^{-1}$. Conversely, if $X$ and a set of conditional probabilities $\left\{   P_{Z|X}(. |x), x\in A^\mathcal{I}\right\}$ (i.e. $F$) are given, then $P_{XZ}$ and the output process $Z$ are well defined. A functional notation $Z=F(X)$ will be used below.

The desired (i.e. perfect) computation will be represented by a {\em measurable} function $f: A^\mathcal{I} \rightarrow B^\mathcal{I}$ where $B$ is a countable alphabet.  $Y=f(X)$ is a random process of  distribution $P_Y=P_Xf^{-1}$. The function $f$ defines a deterministic channel $X \rightarrow Y=f(X)$ which is a set of conditional probabilities $\left\{   P_{f(X)|X}(. |x), x\in A^\mathcal{I}\right\}$ (See \cite{Gray11}, chap. 2): 
\begin{equation}
	\forall G \in \mathcal{B}_{B^\mathcal{I}},  P_{f(X)|X}(G|x) = 1_{f^{-1}(G)}(x) \text{ 	}P_X\text { a.e.}  \label{EquationDeterministicChannel}
\end{equation}

\begin{figure}[htbp]
	\centering
	\xymatrix{
	\left( A^{\mathcal{I}} , \mathcal{B}_{A^\mathcal{I}}, P_X \right) 	\ar @/^1.5pc/ [rr]^f \ar [dr]_F	&						& \left( B^{\mathcal{I}} , \mathcal{B}_{B^\mathcal{I}}, P_Y  \right) \ar @/_-1pc/  [ll]^{f^{-1}} \ar [dl]^{f^{-1}F}\\
 													&  \left( C^{\mathcal{I}} , \mathcal{B}_{C^\mathcal{I}}, P_Z \right)
 	 }
	\caption{Model for Noisy Computation}
	 \label{FigNoisyComputationModel}
\end{figure}
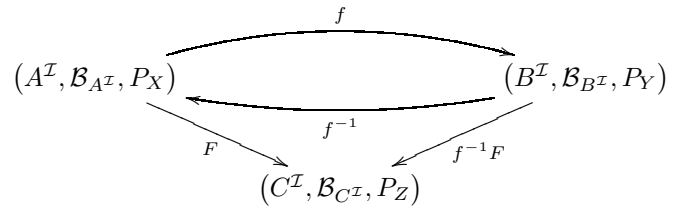

$F$ and $f$ determine a channel $f(X)\rightarrow Z$ which is a cascade of the reverse channel $f^{-1}$ followed by $F$ (figure~(\ref{FigNoisyComputationModel}).
The noisy computation model should not be understood as a cascade made of a perfect function followed by a noisy channel (as done in \cite{WinogradCowan63} for instance). The cascade $f^{-1}F$ is an "artifact" on which  a channel coding theorem will be invoked to build a code for the noisy computation.

From the expression giving the probabilities of a cascade of channels  (\cite{Gray11}, chap. 2) and applying (\ref{EquationDeterministicChannel}), it easily comes that $P_{Y|Z}\equiv P_{X|Z}f^{-1}$.

 If $f$ is bijective and $A=B$, the model is the noisy channel one. 

\section{Capacity of a Noisy Computation \label{SectionCapacityofaNoisyComputation}}

The alphabets are now assumed finite and the $\sigma$-fields $\mathcal{B}_A$, $\mathcal{B}_B$  and $\mathcal{B}_C$ are the sets of subsets of $A$, $B$ and $C$. $f^n(X^n)$  stands for the random value $Y^n$ (the $n$ first symbols of $Y$), $F^n(X^n)$ for $Z^n$. $[A\times B, Xf ; A\times C, XF]$ denotes the noisy computation $F$ of $f$ on $X$. 
{\em $X$, $F$, $f$ and the cascade $f^{-1}F$ are assumed AMS and ergodic}. Thus entropy rates are limits and ergodic theorems hold (\cite{Gray11}).

The Feinstein's theorem is reminded (\cite{Gray11}, chap. 12):
\begin{theorem}[Feinstein's theorem]
Let $[A\times B, \mu\nu]$ be an AMS and ergodic hookup of a source $\mu$ and channel $\nu$. Let $\overline{I}_{\mu\nu}=\overline{I}_{\mu\nu}(X;Y)$ denote the average mutual information rate and assume that $\overline{I}_{\mu\nu}$ is finite (this is the case if the alphabets are finite.). Then for any $R<\overline{I}_{\mu\nu}$ and for any $\epsilon >0$, there exists, for $n$ large enough, a code $\left\{ (\omega_i, \Gamma_i) \in A^n \times \mathcal{B}_{B^n} , i=1,\cdots,M \right\}$
such that $M = \lfloor e^{nR} \rfloor $ and $\forall i=1,\cdots,M,  \hat{\nu}^n(\Gamma_i^c|\omega_i) \leq \epsilon$
\end{theorem}
$\hat{\nu}^n$ is the channel induced by the source $\mu$, (\cite{Gray11}, chap. 12).

The following definition introduces the {\em typical input rate} which will be shown to be the rate at which a source should produce typical inputs for a random process to allow to recover the desired function results by decoding.

\begin{definition}
The {\em typical input rate} of the source $X$ for the noisy computation $F$ with respect to the perfect function $f$ is the following limit, denoted $\overline{B}(X,f,F)$: 
$$
	lim_{n\rightarrow \infty }\frac{ H(X^n|f^n(X^n)) + I(F^n(X^n);f^n(X^n))}{n} \nonumber
$$
\end{definition}
Taking into account the hypothesis ( AMS and ergodic source and channels), $\overline{B}(X,f,F)$  is  well defined. Simple algebra gives $\overline{B}(X,f,F)=\overline{H}(X) - \overline{H}(f(X)|F(X))$

\begin{definition} \label{DefinitionFeinsteinCodeForNoisyComputation}
Let $[A\times B, Xf ; A\times C , XF]$ be a noisy computation on finite alphabets $A$, $B$ and $C$. A {\em $[M,n,\epsilon]$-Feinstein code} for the noisy computation $[A\times B, Xf ; A\times C , XF]$ is a set $\left\{(A^n_i, \Gamma^n_i)\in \mathcal{B}_{A^n}\times \mathcal{B}_{C^n}, i=1,\cdots,M \right\}$ such that:
\begin{enumerate}
	\item $\hat{P}^n_{F(X)|X}(\Gamma_i^c|x^n) \leq \epsilon$ for any $x^n\in A^n_i$, $i=1,\cdots,M$
	\item $\forall i=1,\cdots,M, \exists y_i \in B^n \text{ such that  } A^n_i =  (f^n)^{-1}(y_i)$
\end{enumerate}
$F^n(X^n)$ {\em $\epsilon$-reliably computes} $f^n(X^n)$ on the code $\left\{(A^n_i, \Gamma^n_i), i=1,\cdots,M \right\}$.
\end{definition}

\begin{lemma} [Feinstein's theorem for Noisy Computation] \label{TheoremNoisyComputation}
Let $[A\times B, Xf ; A\times C , XF]$ be a noisy computation on finite alphabets. For any $R<\overline{B}(X,f,F)$, for any $\epsilon>0$, n large enough, there exists a $[\lfloor e^{n(R-\overline{H}(X|f(X)))} \rfloor,n,\epsilon]$-Feinstein code for $F^n(X^n)$ to $\epsilon$-reliably computes $f^n(X^n)$.
\end{lemma}
\begin{IEEEproof}
$R<\overline{B}(X,f,F) \Rightarrow R'=R-\overline{H}(X|f(X)) < \overline{I}(f(X),F(X))$. Then, thanks to the Feinstein's theorem, since $P_{f(X),F(X)}$ is AMS and ergodic (by assumption), for $n$ large enough, there exists a Feinstein code $\{ (y_i, \Gamma_i) \in B^n \times \mathcal{B}_{C^n} ; i=1,\cdots, M\}$ such that $M=\lfloor e^{nR'} \rfloor$ and $\forall i=1,\cdots,M ; \hat{P}^n_{F(X)|X}(\Gamma_i^c|y_i) \leq \epsilon$

Let $x$ belong to $ (f^n)^{-1}(y)$. Considering  the cascade $X^n\rightarrow f^n(X^n)\rightarrow F^n(X^n)$, for any $k=1,\cdots,M$:
$$
\hat{P}^n_{F(X)|X}(\Gamma_k^{c}|x) = \int_{B^{\mathcal{I}}} \hat{P}^n_{F(X)|f(X)}(\Gamma_k^{c}|\underline{y}) d\hat{P}^n_{f(X)|X}(\underline{y}|x)
$$
\begin{multline}
 \hat{P}^n_{F(X)|X}(\Gamma_k^{c}|x)  = 
 	\int_{\left\{y_k \right \}} \hat{P}^n_{F(X)|f(X)}(\Gamma_k^{c}|\underline{y}) d\hat{P}^n_{f(X)|X}(\underline{y}|x) \nonumber \\
				    + \int_{\left\{y_k \right \}^c} \hat{P}^n_{F(X)|f(X)}(\Gamma_k^{c}|\underline{y}) d\hat{P}_{f(X)|X}(\underline{y}|x) \nonumber \\
			         		 \leq   \hat{P}^n_{f(X)|X}\left(  \left\{y_k \right \} |x \right).\epsilon +\hat{P}^n_{f(X)|X}\left(   \left\{y_k \right \}^c |x \right) \nonumber
\end{multline}
If $x\in (f^n)^{-1}(y_k)$ then $\hat{P}^n_{f(X)|X}\left(  \left\{y_k \right \} |x \right)=1$ and $\hat{P}^n_{f(X)|X}\left(  \left\{y_k \right \}^c |x \right)=0$, hence $\forall x\in f^{-1}(y_k)$,  $ \hat{P}^n_{F(X)|X}\left( \Gamma_k^c | x_k  \right) \leq \epsilon$
\end{IEEEproof}

We can conclude this section by the definition of the {\em typical input capacity} of a noisy computation.

\begin{definition}
The typical input capacity of the noisy function $F$ with respect to the perfect function $f$ is $C_f(F) = \sup_{\text{AMS erg  } P_X} \overline{B}(X,f,F)$, the supremum is over all AMS and ergodic sources $X$.
\end{definition}

The equivalent expression $C_f(F) = \sup_{\text{AMS erg  } P_X} \left[ \overline{H}(X) - \overline{H}(f(X)|F(X))\right]$  shows that this capacity boils down to the "usual" channel capacity when $f$ is a bijection, in which case $\overline{H}(f(X)|F(X))=\overline{H}(X|F(X))$.

\section{Reliable computation\label{SectionReliableComputation}}
There is a need (\cite{Elias58,Winograd62}) to constrain the encoding and decoding processes to avoid  the following cases:
\begin{itemize}
	\item either an (assumed perfect) encoder which computes the expected function, encodes the result before transmission through the random process (considered as a noisy transmission channel)
	\item or an encoder which encodes input values for reliable transmission through the random process (considered here also as a noisy transmission channel) and a decoder (assumed also reliable) which decodes (almost perfectly) and computes (perfectly) the expected function.
\end{itemize}

Considering that a computation $g$  is a "true" computation if the entropy is reduced ($\overline{H}(X'|g(X'))>0 \Rightarrow \overline{H}(g(X')) < \overline{H}(X')$ (else it is communication), the model must be targeted to be  mainly relevant for non-injective functions (i.e. $\overline{H}(X'|g(X'))>0$). For injective $g$, this becomes  the classical reliable transmission model. With the constraint that both encoding and decoding are based on injections (in a sense made precise below) then the encoder and the decoder cannot compute (at least totally) the desired function as they do not reduce entropy. 

The proposed model of the complete process to reliably compute a  function $g :A'^{\mathcal{I}} \rightarrow B'^{\mathcal{I}}$ acting on a source $X'$, thanks to a  noisy implementation $F$ of a  function $f: A^{\mathcal{I}} \rightarrow C^{\mathcal{I}}$ is the following:
\begin{itemize}
	\item {\bf encoding}: let $X^n$ be the $n^{th}$ extension of an  source for which we have a maximal code $(A^n_i,\Gamma_i)_{i=1,\cdots,M}$ allowing to $\epsilon$-reliably compute $f^n(X^n)$ by $F^n(X^n)$ (cf lemma~\ref{TheoremNoisyComputation} and definition~\ref{DefinitionFeinsteinCodeForNoisyComputation}) ; a typical $k$-sequence $x'$ of $X'^k$ is encoded into a {\em typical given $y_i$} (this important assumption is discussed in the conclusion) $n$-sequence of $X^n$ by a injective function, say $\mathcal{U}$, such that $\mathcal{U}(x') \in A_i^n$ for some $i=1,\cdots,M$
	\item {\bf computation of the noisy function}: $F^n$ is applied to $\mathcal{U}(x')$ producing a typical $n$-sequence $F^n(\mathcal{U}(x'))$ of $F^n(X^n)$ where $F^n(\mathcal{U}(x'))$ belongs to a given $\Gamma_i$ (with probability greater than $1-\epsilon$)
	\item {\bf decoding}: the first step is to associate to $F^n(\mathcal{U}(x'))$ the typical $n$-sequence $y_i$ of $f^n(X^n)$ corresponding to $\Gamma_i$, the second step is to apply to $y_i$ a function $\mathcal{V} : \{\mathbf{y_1},\ldots,\mathbf{y_M}\} \rightarrow \{\text{typical k-sequences of }g^k(X'^k)\} $ such that $\mathcal{V}(y_i)=g^k(x')$
\end{itemize}
A decoding error occurs when one obtains a $n$-sequence $y_j$ (or equivalently a  $\Gamma_j$) such that $\widehat{g^k(x')} =  \mathcal{V}(y_j) \neq g^k(x')$

To be able to define a decoding function $\mathcal{V}$ (i.e, a deterministic decoding), the encoding function $\mathcal{U}$ has to be such that the typical (given $y_i$) $n$-sequences of one $A_i^n = (f^n)^{-1} (y_i)$ ($ y_i \in \{y_1,\ldots,y_M\}$) are used for encoding typical $k$-sequences of {\em only one} $(g^k)^{-1} (z)$, $z$ typical $k$-sequence of $g^k(X'^k)$.

We also require that $\mathcal{V}$ be an injection (as we have required from $\mathcal{U}$). 

The typical $k$-sequences of a $(g^k)^{-1} (z)$, $z$ typical $k$-sequence of $g^k(X'^k)$, are encoded in typical (given $y_i$) $n$-sequences of one and only one $A_i^n$ . So, if $x'_1$ and $x'_2$ are two typical $k$-sequences of $X'^k$: 
$$
f^n(\mathcal{U}(x'_1)) = f^n(\mathcal{U}(x'_2)) \Leftrightarrow g^k(x'_1) = g^k(x'_2)
$$

The model fulfills the constraints identified above. The encoder implements an injection and thus cannot compute the desired function $f$ nor $g$ (if $f$ and $g$ are not injective). The same comment applies to the injective decoding step  $\mathcal{V}$. 

\section{A coding theorem and its converse\label{SectionCodingTheorem}}

The sources, functions, noisy function and the cascade $f^{-1}F$ are assumed AMS and ergodic.

\begin{definition}   \label{DefinitionEncodingRate}
With the notations of section \ref{SectionReliableComputation}, the ratio $R=\frac{k.H(X')}{n}$ is called the {\em typical encoding input rate}. A rate $R$ is said to be {\em achievable with respect to the function $f$} if there exists a sequence of codes of size $n$ such that the maximal probability of decoding error tends to 0 as $n$ tends to infinity.
\end{definition}

\begin{theorem} \label{TheoremCoding}
If $R < C_f(F)$, then $R$ is achievable w.r.t $f$. 
\end{theorem}

\begin{IEEEproof}

This proof, although identical to that in \cite{Simon10}, is given as it includes the starting point for the proof of the converse theorem. 
 First, it is shown that the injective encoding of typical $k$-sequences of a set $(g^k)^{-1} (z)$ on typical (given $y_i$) $n$-sequences belonging to $A_i^n$ is possible for suitably chosen $k$ and $n$ (lossless coding). Secondly, it is shown that, at encoding input rates below capacity and for $k$ and $n$ suitably chosen,  the sets $A_i^n$ are almost as many as the sets $(g^k)^{-1} (z)$.

Let $\delta'' >0$. Since $\mathbb{Q}$ is dense in $\mathbb{R}$, there exist $k$ and $n$ such that:
$$
\frac{\overline{H}(X'|g(X'))}{\overline{H}(X|f(X))} < \frac{n}{k} <\frac{\overline{H}(X'|g(X'))+\delta''}{\overline{H}(X|f(X))}
$$
Moreover, $k$ and $n$ can be chosen as large as needed. Thus:
\begin{equation}
\frac{k.\overline{H}(X'|g('X))}{H(X|f(X))} < n <\frac{k.(\overline{H}(X'|g(X'))+\delta'')}{\overline{H}(X|f(X))}  \label{InequalitySourceCoding}
\end{equation}
We can choose $\delta, \delta' >0$ and $0<\epsilon<1/2$ small enough for:
\begin{multline*}
	\frac{k.\left( \overline{H}(X'|g(X'))+\delta \right) - log(1-2\epsilon)}{\overline{H}(X|f(X))-\delta'} < n \\
	<\frac{k.(\overline{H}(X'|g(X'))+\delta+\delta'')}{\overline{H}(X|f(X))+\delta'}
\end{multline*}
giving
\begin{multline*}
	k.(\overline{H}(X'|g(X'))+\delta)  < log(1-2\epsilon) + n.(\overline{H}(X|f(X))-\delta') \\
 	< n.(H(X|f(X))+\delta') < k.(H(X'|g(X'))+\delta+\delta'')
\end{multline*}
If $\nu_1$ is the number of typical $k$-sequences of $(g^k)^{-1} (z)$ and $\nu_2$ is the number of typical (given $y_i$) $n$-sequences in an $A_i^n$, we have (by conditional AEP):
\begin{multline*}
	\nu_1 < e^{k.(\overline{H}(X'|g(X'))+\delta)}  <(1-2\epsilon)e^{ n.(\overline{H}(X|f(X))-\delta')} \\
	<\nu_2 <e^{ n.(\overline{H}(X|f(X))+\delta')} <e^{ k.(\overline{H}(X'|g(X'))+\delta+\delta'')}
\end{multline*}
It is thus possible to find an injection from the set of typical $k$-sequences of $(g^k)^{-1} (z)$ on the subset of typical sequences (given $y_i$) of $A_i^n$. This shows the first step. 

Assume that $R=k\overline{H}(X')/n< \overline{H}(X)-\overline{H}(f(X)|F(X)) \leq C_f(F)$. Such a  $X$ exists by definition of $C_f(F)$.  So
\begin{multline*}
	k(\overline{H}(g(X'))+\overline{H}(X'|g(X'))) < \\
 	 n.(\overline{H}(f(X))-\overline{H}(f(X)|F(X)))+n.\overline{H}(X|f(X))
\end{multline*}

By (\ref{InequalitySourceCoding}), $n.\overline{H}(X|f(X)) - k.\overline{H}(X'|g(X')) < k.\delta''$
thus 
$$
k\overline{H}(g(X')) < n.(\overline{H}(f(X))-\overline{H}(f(X)|F(X)))+ k.\delta''
$$
$\epsilon_1,\delta''' >0$ can be chosen small enough in order to get:
$$
e^{k(\overline{H}(g(X'))+\delta''')} < e^{n.(\overline{H}(f(X))-\overline{H}(f(X)|F(X))+ \frac{k}{n}.\delta''-\epsilon_1)}
$$

If $\nu_3$ is the number of typical $k$-sequences of $g^k (X'^k)$ and $M$ is the size of the code (i.e., the number of $(A_i^n,\Gamma_i)$), we have (by AEP and Lemma~\ref{TheoremNoisyComputation}):
\begin{multline*}
\nu_3 < e^{k(\overline{H}(g(X'))+\delta''')} \\ 
	< e^{n.(\overline{H}(f(X))-\overline{H}(f(X)|F(X))+ \frac{k}{n}.\delta''-\epsilon_1)} <M
\end{multline*}
\end{IEEEproof}

The assumed model, by the constraints on encoding, implies that the best ratio (i.e., the smaller) $\frac{n}{k}$ of encoding respects the inequality~(\ref {InequalitySourceCoding}):  $\frac{\overline{H}(X'|g(X'))}{\overline{H}(X|f(X))} < \frac{n}{k}<\frac{\overline{H}(X'|g(X'))+\delta''}{\overline{H}(X|f(X))}$. Let $\gamma=\frac{\overline{H}(X'|g(X'))}{\overline{H}(X|f(X))}$. To respect the encoding constraints (typical sequences are "injectively" encoded into typical sequences), a rate $R=\frac{k}{n}\overline{H}(X')$ must be such that $R\leq \frac{\overline{H}(X')}{\gamma}$

\begin{theorem} \label{TheoremConverse}
If $R>C_f(F)$, there is no code such that the error probability tends to $0$ as $n \rightarrow \infty$
\end{theorem}

\begin{IEEEproof}
The decoding is deterministic then:
\begin{enumerate}
	\item $F^n(X^n)\rightarrow f^n(X^n)  \rightarrow g^k(X'^k)$ is a Markov Chain  thus $g^k(X'^k)\rightarrow f^n(X^n)  \rightarrow F^n(X^n)$ is a Markov Chain
	\item $f^n(X^n)\rightarrow F^n(X^n)  \rightarrow \widehat{g^k(X'^k)}$ is a Markov Chain
\end{enumerate}
Hence $g^k(X'^k)\rightarrow f^n(X^n)  \rightarrow F^n(X^n)  \rightarrow \widehat{g^k(X'^k)} $ is a Markov Chain. This implies that, $\forall n,k \text{  such that }\frac{k}{n} \leq \frac{1}{\gamma}$:
\begin{multline}
I(g^k(X'^k) ; \widehat{g^k(X'^k)})  \leq  I(f^n(X^n) ; F^n(X^n)) \text{ hence}\nonumber \\
 H(g^k(X'^k)) - H(g^k(X'^k)|\widehat{g^k(X'^k)})  \leq  \nonumber \\
                                           			   I(f^n(X^n) ; F^n(X^n)) \nonumber \\
\Rightarrow H(X'^k)- H(X'^k | g^k(X'^k)) - H(g^k(X'^k)|\widehat{g^k(X'^k)})  \leq \nonumber \\
                I(f^n(X^n) ; F^n(X^n)) \nonumber
\end{multline}
by Fano's inequality:
\begin{multline}
H(X'^k)- H(X'^k | g^k(X'^k)) \nonumber \\
	- (H_2(P_e(k))+k.P_e(k).log(|B'|))  \nonumber \\
		\leq  I(f^n(X^n) ; F^n(X^n)) \nonumber \\
\Rightarrow \frac{H(X'^k)}{k}  - \frac{H_2(P_e(k))}{k} - P_e(k).log(|B'|)  \leq  \nonumber \\
	\frac{H(X'^k | g^k(X'^k))}{k} + \frac{ I(f^n(X^n) ; F^n(X^n))}{k}\nonumber \\
\Rightarrow \frac{H(X'^k)}{k}  - \frac{H_2(P_e(k))}{k} - P_e(k).log(|B'|)  \leq \nonumber \\
	 \frac{H(X'^k | g^k(X'^k))}{k} + \gamma . \frac{ I(f^n(X^n) ; F^n(X^n))}{n}\nonumber 
\end{multline}

If the error probability is asymptotically $0$ (i.e., $lim_{k \rightarrow \infty}(P_e(k)=0$) then necessarily (letting $k$ and $n$ tend to infinity): $\overline{H}(X')  \leq  \overline{H}(X' | g(X')) + \gamma . \overline{I}(f(X) ; F(X))$. 
But $\overline{H}(X' | g(X')) = \gamma \overline{H}(X | f(X))$, then:
$$
\overline{H}(X')  \leq  \gamma .\left( \overline{H}(X | f(X)) + \overline{I}(f(X) ; F(X)) \right)  
$$
since $R\leq \frac{\overline{H}(X')}{\gamma}$, we obtain
$$
R \leq \overline{H}(X | f(X)) + \overline{I}(f(X) ; F(X)) \leq C_f(F)
$$
Thus if $R > C_f(F)$ then the error probability does not vanish.
\end{IEEEproof}

\section{Discussion and conclusion \label{SectionConclusion}}

The coding lemma, the coding theorem and its converse assume that the sources ($X$ and $X'$),  the channels ($F$, $f$, $g$) and the cascade $f^{-1}F$ are AMS and ergodic. Cases can be identified where it is possible to derive the AMS property and ergodicity of  $f^{-1}F$ from properties of $X$, $F$ and $f$ (e.g., if $X$, $F$ are stationary and weakly mixing and $f$ AMS and ergodic then $f^{-1}F$ is AMS and ergodic). Due to lack of space, this question is not addressed here, neither the identification of classes of AMS and ergodic functions $f$ and $g$.

The model of reliable computation assumes that the encoder and the decoder are perfectly reliable. This assumption could be justified by quoting from \cite{Winograd62} {\em "The computation system [model] was devised for the sole purpose of studying the relation of information theory of reliable automata"}. Moreover we could argue in addition that if the complexity of the computation device is of a much greater magnitude than that of the encoder and decoder then the unreliability of the encoder and decoder have almost no impact on the overall reliability of the computation and thus can be neglected. For complex systems (e.g., based on significant software volume), this is quite realistic. In any case, it is impossible to overcome the fact that the reliability reached is at the best the reliability of the final decoding device. The only way is to built a intrinsically reliable enough decoder (for example thanks to gate redundancy).  A noisy encoder is a noisy computation itself and thus can be handled from the point of view of "cascaded noisy computations". This is outside of the scope of the present paper.

The proposed model of reliable computation involves two perfect functions $ g$ and $f$. This is intended to capture major real cases as already mentioned. Another motivation to use an "ancillary" function $f$ in the model is that this is an efficient way to define an input code,  meaning a family of subsets $(A^n_i)_{i=1,\cdots,M}$, that do not overlap and whose "images" by the noisy function do not overlap "too much" (i.e., fall into disjoint $\Gamma_i$ with high probability). Defining such family is defining (partially) a function $f$ by picking, for each $i$ an $y_i$ and stating $f^{-1}(y_i)=A_i$. In addition $f$ allows a characterization of a kind of size of the sets $(f^n)^{-1}(y_i)$ through the conditional entropy rate $\overline{H}(X|f(X))$. This motivates also  the constraint of coding by conditionally typical sequences. While the sets $(f^n)^{-1}(y_i)$ are "balanced", for large $n$, with respect to the number of (conditionally) typical sequences they contain, their cardinalities might be very different and bounds are not straightforward to obtain. Thus, the use of all possible  elements of $(f^n)^{-1}(y_i)$ forbids to characterize all the $(f^n)^{-1}(y_i)$ by the same number measuring the "encoding" power. The same difficulty forbids to state a converse as well. The "encoding by conditionally typical sequences" trick overcomes this difficulty.

\bibliographystyle{IEEEtran}
\bibliography{IEEEabrv,NoisyComputations}

\end{document}